\documentclass[sn-mathphys,Numbered]{sn-jnl}

\usepackage{graphicx}%
\usepackage{multirow}%
\usepackage{amsmath,amssymb,amsfonts}%
\usepackage{amsthm}%
\usepackage{mathrsfs}%
\usepackage[title]{appendix}%
\usepackage{xcolor}%
\usepackage{textcomp}%
\usepackage{manyfoot}%
\usepackage{booktabs}%
\usepackage{algorithm}%
\usepackage{algorithmicx}%
\usepackage{algpseudocode}%
\usepackage{listings}%
\usepackage{tabularx}%

\theoremstyle{thmstyleone}%
\theoremstyle{thmstyletwo}%

\theoremstyle{thmstylethree}%

\begin{document}

\title[Article Title]{Unsupervised Learning of Molecular Embeddings
for Enhanced Clustering and Emergent Properties
for Chemical Compounds}


\author{Jaiveer Gill$^*$}\email{jaiveer.gill26@bcp.org}
\author{Ratul Chakraborty$^*$}\email{imratulc@gmail.com}
\author{Reetham Gubba}\email{gubba.ricky@gmail.com}
\author{Amy Liu}\email{amyliu123005@gmail.com}
\author{Shrey Jain}\email{shreyraij@gmail.com}
\author{Chirag Iyer}\email{chiragkarthikiyer@gmail.com}
\author{Obaid Khwaja}\email{obaidzkhwaja@gmail.com}
\author{Saurav Kumar}\email{sauravk4@illinois.edu}





\abstract{

Molecular structure and property analysis plays a pivotal role in drug development, with significant potential for advancement through machine learning. The development of an emergent property model to decipher molecular intricacies offers a novel computational approach. In this study, we introduce methods for the detection and clustering of chemical compounds based on SMILES data. We designed and developed a similarity search algorithm that uses the Tanimoto coefficient and molecular fingerprinting to analyze graphical chemical structures. Additionally, we enhance existing LLMs using natural language description embeddings stored in a vector database. Our efficient similarity search and clustering algorithm resulted in distinct clusters exceeding a predefined threshold. This approach has the potential to pave the way for transformer-based drug design, offering researchers deeper insights into molecular properties through autoencoders and similarity search.

}



\maketitle

\section{Introduction}\label{sec1}
Despite a considerable amount of research towards synthesizing specific molecules, the fields of biochemistry and pharmacology still lack a machine learning approach that enables understanding of underlying properties in molecules. Such an approach would benefit various fields by allowing intelligent querying for molecules based on their properties and maintaining sustainability in developing novel solutions in future drug discovery, and can rapidly advance how future researchers interact with molecular information. Natural language processing and embeddings can be used on both raw text data as well as graphical molecular data. This makes the creation of medical-grade drugs and chemicals not only easier but also more transparent. 

Natural language queries are characterized by inputs that are entered as spoken or written language, but lack special and punctuation characters such as the ‘+’ or ‘!’. These are processed by a large language model, using embeddings and large language models A variety of embedding techniques are used to transform the input text into a series of numbers for processing. 

Current molecule searching and drug discovery methods do not give researchers the outright ability to analyze molecules using a reasoning methodology or to infer the properties of novel drugs. Instead, these current functions are limited in their capacity and not intuitive, relying on data that are not readily comprehensible, such as SMILES (Simplified Molecular Input Line Entry System), for marking specific queries, which are very complex and long strings of data unable to be consistently and thoroughly analyzed by most. As a result, essential similarities and properties for drug discovery are hidden within the complex formatting in both the graphical representation of these molecules as well as the natural language descriptions. Our aim through these methods is to develop models capable of developing emergent properties that enable machines to understand molecular properties and relations through either graphical data or descriptions, giving researchers easy access to certain properties and tools, and increasing understanding of molecules and making drug discovery a much faster process as a result. 

Furthermore, in Rives et al, the authors use a transformer neural network to process the structures and functions of 86 billion amino acids and later use embeddings to represent protein sequences as points in higher dimensional spaces. Using a self-supervised model to apply protein data as unlabeled amino acid sequences allowed this transformer model to provide a characterization of the proteins and amino acids–the model could have representations that contain information about the biological properties of amino acids and proteins. This method created a deep language contextual model to advance the predictive and generative capabilities of artificial intelligence in biology. Our project builds upon this method, exploring its generative and predictive applications in biochemistry with chemical property prediction. 

In Bran et al, the authors discuss “Augmenting large-language models with chemistry tools,” and introduce a novel chemistry-based NLP model called ChemChrow. The authors in this text outline their ChemChrow model, which is an LLM-based chemistry agent meant to answer questions about chemistry-based queries using information from web scraped data as needed regarding the query. Their approach created a model based on GPT to aid in drug discovery and to help less knowledgeable researchers delve into the world of chemistry. Our project explores using this method to generate descriptions for each given molecule in our dataset to make the data richer to train our model against. In future research, we would also delve into using ChemChrow as a benchmark to evaluate our model’s understanding of properties in molecules. 

In Jumper et al, the paper "Highly Accurate Protein Structure Prediction with AlphaFold" published in Nature presents a transformative advancement in the field of molecular biology. The AlphaFold algorithm, developed by researchers at DeepMind, utilizes a deep-learning approach to predict protein structures with unprecedented accuracy. By leveraging a variant of the attention mechanism within transformer models, AlphaFold demonstrates remarkable proficiency in predicting three-dimensional protein structures. This innovation holds immense significance within our broader application of detecting properties in molecules. As the accurate prediction of protein structures is a pivotal step in understanding molecular behavior and function, the insights and methodology derived from AlphaFold's success could potentially serve as a cornerstone for enhancing our ability to develop emergent properties to understand various molecules.

\section{Results}
Our study introduced important methods and steps toward the advancement of research in molecular property prediction and the goal of developing emergent properties in models. Our similarity search algorithm, which employs vector search, was able to accurately identify and cluster chemical compounds based on their graphical input data. This algorithm can be used in the future in parallel with other transformer architectures as a molecular property prediction tool, potentially using a textual input or output to enrich the model. This would employ transformer architecture that can understand fundamental properties in molecules while still being able to understand textual relations. Our vector database of NLP embeddings or fine-tuned models can aid this process to potentially revolutionize the field of drug development, overall allowing researchers to more easily understand the underlying properties of molecules and develop novel solutions in future drug discovery.

One of the key impacts of our study is its potential to significantly improve the efficiency and accuracy of chemical exploration. By leveraging our similarity search algorithm and natural language description embeddings stored in a vector database with GPT-3.5, researchers can more easily search for molecules based on natural language queries and structural similarity. This can help to maintain sustainability in developing novel solutions for future drug discovery, and can rapidly advance how future researchers interact with molecular information.

Our algorithm for similarity search for molecules has the potential to be a powerful tool for scientists to research and develop new drugs and chemicals. By accurately identifying and clustering chemical compounds based on their SMILES data, our algorithm can help researchers to more easily understand the underlying properties of molecules and develop novel solutions in future drug discovery. By giving transformers an understandings of how structural similarities work, we also move closer to a transformer that understands the impact of structural differences in molecules. This can lead to the development of more effective and efficient drugs and chemicals, ultimately benefiting society as a whole and paving the way for future developments in drug discovery.

\section{Natural Language Data Collection}
While extensive chemical compound information is available online through the PubChem dataset, it lacks descriptions suitable for robust training data for a large language model due to variations in length and information richness. To address this, we leveraged the PUG-View API, a REST-style web service provided by PubChem, to aggregate descriptions for each molecule in the dataset.

This effort resulted in the creation of a dataset comprising 328,000 compound descriptions, including SMILES data, chemical formulas, molecular weights, and other pertinent information for each compound. Although the dataset drawn solely from compounds with descriptions provided a reasonable size for research purposes, the average length of these descriptions was a mere 3.09 sentences, leaving room for improvement in fine-tuning.

Our initial attempt at enhancing the dataset involved developing a web scraper to extract relevant information about target molecules from reputable scientific journals to enrich the descriptions. Instead, we employed alternative methods to enhance the dataset and maximize potential accuracy scores for the final model. One such method involved concatenating the 'description' data with all other data categories in the dataset, such as molecular weight, molecular formula, and polar area. This resulted in an increase in the average length of each description by incorporating relevant information, while maintaining the richness of the descriptions. Additionally, the inclusion of molecular formula, polar area, and molecular weight contributed to enhancing the variability and uniqueness of each description in the dataset.

 \begin{table}[ht]
    \centering
    \vspace{10pt} 
    \centering
    \caption{Original Dataset Response}
    \begin{tabular}{p{0.4\linewidth}p{0.5\linewidth}}
        \toprule
        \textbf{Prompt} & \textbf{Response} \\
        \midrule
        Ten heaviest gases? & No info on ten heaviest gases. \\
        \bottomrule
    \end{tabular}

\end{table}

\begin{table}[ht]
    \centering
    \small 
    \caption{Response to the prompt with the enhanced dataset}
    \begin{tabular}{p{0.2\linewidth}p{0.25\linewidth}p{0.25\linewidth}p{0.2\linewidth}} 
        \toprule
        \textbf{No.} & \textbf{Gas Name} & \textbf{Weight Type} & \textbf{Weight Value} \\
        \midrule
        1 & Radon (Rn) & Atomic weight & 222.0 \\
        2 & Uranium Hexafluoride (UF6) & Molecular weight & 352.0 \\
        3 & Tungsten Hexafluoride (WF6) & Molecular weight & 297.8 \\
        4 & Sulfur Hexafluoride (SF6) & Molecular weight & 146.1 \\
        5 & Radium (Ra) & Atomic weight & 226.0 \\
        6 & Plutonium Hexafluoride (PuF6) & Molecular weight & 329.0 \\
        7 & Osmium Tetroxide (OsO4) & Molecular weight & 254.2 \\
        8 & Oganesson (Og) & Atomic weight & 294.0 Heaviest element) \\
        9 & Nitrous Oxide (N2O) & Molecular weight & 44.0 \\
        10 & Xenon (Xe) & Atomic weight & 131.3 \\
        \bottomrule
    \end{tabular}
\end{table}

\section{Graph Neural Network}
Further enhancement of our data took place by adding blood-brain barrier permeability as a data point in our natural language description.

In the context of molecular representation, the GNN takes in a SMILES string and turns it into a graph where the nodes represent atoms and the edges represent bonds. Fig 1 shows the SMILES string of a caffeine molecule turned into an adjacency matrix. Every atom (node) has a feature vector, which represents the attributes of the atom. The GNN iteratively updates the feature vectors of nodes by aggregating information from their neighbors, known as message passing. 

We used a GNN to enhance the dataset by increasing descriptions and as a benchmark for blood-brain barrier property prediction performance in the LLM, which will be discussed later. The blood-brain barrier (BBB) is a highly selective and semi-permeable boundary that separates the circulating blood from the brain and its surrounding extracellular fluid, functioning as a protective barrier from potentially harmful substances while allowing essential nutrients and molecules to pass through. Thus, the controlled permeability of the BBB plays a pivotal role in maintaining brain health, and the permeability property of certain molecules is vital information in developing drugs meant to target the central nervous system.

To predict Blood Brain Barrier Permeability, we utilized the message-passing neural network (MPNN) model detailed by Keras. The MPNN is trained on the benchmark dataset developed by MoleculeNet, and it contains 2050 molecules that each come with a name, label, and SMILES string. The model reported a 96.28 percent AUC (Area Under ROC curve) after training and a 90.26 percent validation AUC after testing. We took the PubChem dataset and ran the SMILES string column through the MPNN to get permeability predictions for all 328k molecules. Because the MPPNN outputs a float value between 0 and 1, we converted these values into natural language descriptions by running the output data frame through an algorithm that assesses the value of the float and returns the permeability as a string sentence. These sentences were then concatenated to the molecules’ respective descriptions already in the data frame, and this added more data to our dataset. Fig 1 visualizes a representation of the caffeine molecule as an adjacency matrix to graphically characterize the data for the GNN, a format easier for computation.
\begin{figure}[htbp]
    \centering
    \includegraphics[width=\linewidth]{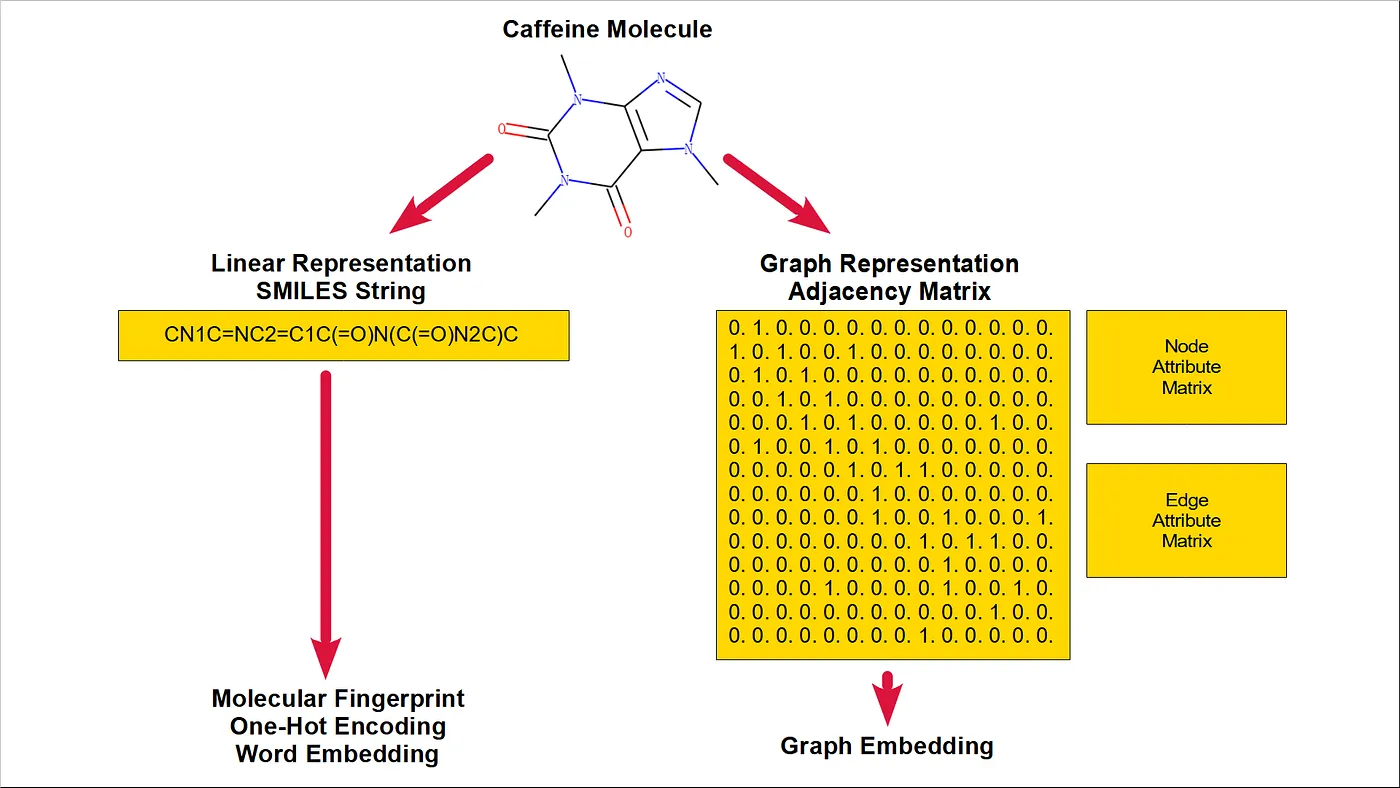}
    \caption{Example graph representation of molecular structure (caffeine)}
    \label{GNN_Mol_Graph_Png.png}
\end{figure}

\section{Fine Tuning}
Using LLM fine-tuning, we were able to analyze the ability of language-based models to characterize molecules. We discuss our methods involving fine-tuning LLMs, namely, LLaMA 2 and OpenAI. Our study aims to use textual data as the foundational training material for the finely tuned model. This augmentation of data plays a pivotal role in fine-tuning, as relevant data is very important to a successfully fine-tuned model. We curated a chemical dataset of 50,000 molecular descriptions, selected based on "richness" or quality of description for the LLM. We used the Hugging Face Transformers library along with the LLaMA 2 model and OpenAI's 'curie' model. The goal of this approach was to experiment with whether purely text-based large language models could form emergent properties regarding molecular relationships and structures.

\subsection{Fine-Tuning Parameters}

For LLaMA 2, we leveraged the SFTTrainer component from the Text-to-Text Transfer Learning (TRL) library. We also used Parameter Efficient Fine-Tuning (PEFT) and Low Rank Adaptation (LoRA) in the fine-tuning process to lower the computational burden. This approach achieves performance comparable to full fine-tuning, while also significantly reducing GPU memory requirements. The application of PEFT and LoRA enhances the computational efficiency of LLaMA fine-tuning, making multiple iterations of fine-tuning a feasible avenue for optimal results. Curie's API-based fine-tuning method led to much less direct involvement with hyperparameters due to it being streamlined by OpenAI.

\subsection{Fine-Tuning Methods}

We utilized several different methods of fine-tuning, all using LLaMA 2 7B and GPT-3 'curie' applied on many unique datasets. The first iteration involved tuning LLaMA 2 on the data generated earlier using PubChem descriptions and properties. The entire dataset was not fine-tuned as most molecules and functional groups relevant to the model's understanding of chemistry lie in the first portion of the dataset. 
\begin{wraptable}{l}{0.4\textwidth}
\centering
\begin{tabular}{|c|c|}
\hline
\textbf{Rank} & \textbf{"Liquid"} \\
\hline
1 & Mercury (Hg) \\
\hline
2 & Bromine (Br) \\
\hline
3 & Iodine (I) \\
\hline
4 & Chlorine (Cl) \\
\hline
5 & Sulfur (S) \\
\hline
6 & Phosphorus (P) \\
\hline
7 & Arsenic (As) \\
\hline
8 & Antimony (Sb) \\
\hline
9 & Bismuth (Bi) \\
\hline
10 & Tellurium (Te) \\
\hline
\end{tabular}
\end{wraptable}
Only the first 20,000 molecules were included in the fine-tuned training data. After ~24 hours of training, the LLaMA model proved its understanding of molecular properties and related information. The model was able to understand and respond to example prompts such as 'What are the ten heaviest liquids' and draw from training data to make an educated answer. However, as noted from the table to the left, this approach contained some inaccuracies despite in other areas showing promise in generalization and understanding of molecular relations. However, in the future, more heavily curated data would lead to even better results capable of making a powerful natural language query-based search method for molecules. Furthermore, curie's increased number of parameters paired with further techniques can improve these results.

\subsection{Molecular LLM}

The more important capability of fine-tuning would be creating an LLM capable of understanding the fundamental properties of what makes up a molecule, or a 'Molecular LLM.' This would result in it being capable of understanding the fundamental graphical structure and inter-relationships of a given molecule. To do this the model would not take any textual description as input, which in general goes against the purpose of an LLM. This would challenge large language models like LLaMA and curie to understand not textual patterns in the data but patterns in the molecular structure of the molecules. The first method acted simply as a base test and involved simply inputting the SMILES string into the LLM as the prompt for the LLM with the goal being the model predicting the Blood Brain Barrier Permeability of the molecule and competing with the GNN discussed earlier/ The goal was for the textually based transformer to recognize patterns in the molecules and bonds present in SMILES data and find a relationship between that and the target BBBP. However, because of the complex nature of the SMILES strings being very intricate and oftentimes long, as expected the LLMs (both curie and LLaMA) were unable to grasp the molecular intricacies and patterns in the SMILES data.

\subsection{NLP SMILES input}

\begin{table}[htbp]
    \centering
    \setlength{\tabcolsep}{8pt} 
    \begin{tabularx}{\linewidth}{|c|X|}
        \hline
        SMILES & CC[N+](C)(C)CC1=CC=CC=C1Br \\
        \hline
        Description & Get a description of the molecule with 13 atoms and 13 heavy atoms given SMILES \\
        \hline
        Atom 1 & C (valency 1) forms 1 single bond with C (valency contribution: 1.0) \\
        \hline
        Atom 2 & C (valency 2) forms 2 single bonds with C (valency contribution: 1.0) \\
        \hline
        Atom 3 & N (valency 4) forms 1 single bond with C and 3 single bonds with N (valency contribution: 1.0) \\
        \hline
        Atom 4 & C (valency 1) forms 1 single bond with N (valency contribution: 1.0) \\
        \hline
        Atom 5 & C (valency 1) forms 1 single bond with N (valency contribution: 1.0) \\
        \hline
        Atom 6 & C (valency 2) forms 1 single bond with N and 1 single bond with C (valency contribution: 1.0) \\
        \hline
        Atom 7 & C (valency 4) forms 1 single bond with C and 2 aromatic bonds with C (valency contribution: 1.5) \\
        \hline
        Atom 8 & C (valency 3) forms 2 aromatic bonds with C (valency contribution: 1.5) \\
        \hline
        Atom 9 & C (valency 3) forms 2 aromatic bonds with C (valency contribution: 1.5) \\
        \hline
        Atom 10 & C (valency 3) forms 2 aromatic bonds with C (valency contribution: 1.5) \\
        \hline
        Atom 11 & C (valency 3) forms 2 aromatic bonds with C (valency contribution: 1.5) \\
        \hline
        Atom 12 & C (valency 4) forms 2 aromatic bonds with C and 1 single bond with C (valency contribution: 1.0) \\
        \hline
        Atom 13 & Br (valency 1) forms 1 single bond with C (valency contribution: 1.0) \\
        \hline
    \end{tabularx}
\end{table}
We converted the SMILES data to graphical data via the RDKit library and represented the output graphs in natural language form. The bond data and atomic data were extracted from the graphical representation and converted into a natural language prompt, displaying each of the bonds in the molecule and their respective atoms and valency contributions. The following table shows an example SMILES NLP representation given the SMILES string input 'CC[N+](C)(C)CC1=CC=CC=C1Br', or the compound known as 'Bretylium,' an antiarrhythmic agent and norepinephrine release inhibitor. This was inputted into Curie, and initially, it yielded a result of ~50 percent, which is equivalent to guessing and an accuracy that was not statistically relevant. However, this was using output data of the entire molecular description, including categories like molecular formula, molecular weight, polar area, etc. Although the model performed rather well on those categories, it was more focused on those and unable to grasp relationships between the input data and the Blood-Brain Barrier Penetration, a more complex property. The model was retrained on the same input data with the output data being strictly limited to BBBP. The model was subsequently run on a benchmark dataset of 2050 molecules referenced earlier in the GNN section. It resulted in 71 percent accuracy, which was a notable improvement from the previous iteration. This proved that the LLM understood some parts of the molecular structure of a given molecule and its effects on BBBP. However, an issue in this prediction data is it being heavily skewed toward one column of the data causing a level of bias toward predicting a positive result. This is not ideal because the goal is an even split in outputs.

\subsection{Adjacency Matrix}

Another attempt was to input an adjacency matrix as the prompt for curie with the completion being BBBP. An adjacency matrix is a mathematical representation used primarily in graph theory to describe the relationships between nodes in a graph. Adjacency matrices are used in various graphical applications, in this case regarding molecules. They provide a convenient and structured way to represent the relationships within a graph, making it easier to analyze and manipulate graph-based data. In this context adjacency matrices were used to convert the graphical representation of molecules with atoms and bonds being nodes and edges into the adjacency matrix. This was inputted as a string into the LLM to see if it could understand this direct graphical data. The output result was far from ideal, and it proved that directly graph-based data represented in forms such as the adjacency matrix are unlikely to be directly decipherable by a large language model like curie or LLaMA. An adjacency matrix would simply be a matrix of generally smaller integers representing the given molecule. Instead, future attempts at creating this type of “molecular LLM” would include using a network based on graph transformers to calculate relationships between graphical data more efficiently than in regular LLMs, but still relate that to textual inputs or outputs. This ability to relate molecular and textual data would be greatly beneficial in the field of drug discovery.

\subsection{'Curie' Fine-Tune}

GPT 3's 'curie' model was fine-tuned on the PubChem descriptions from the final concatenated dataset including molecular formula, molecular weight, polar area, hydrogen bond donors, and blood-brain barrier permeability. Because curie contains nearly twice the number of parameters as LLaMA 2 7B, the fine-tuned model was expected to be deeper and more able to pick up intricacies in the training data. Though the results didn't yet compare with the GPT-3 model using a vector database of embeddings as context (a method discussed more thoroughly later), it proved its basic understanding of molecular properties and their relation to each other in similar compound groups. With more refined data in the future, the model can be improved in the future. Using the current data it is not as accurate but with a scaled approach and more heavily curated data it can become a great method for understanding molecular properties and their relations.

\section{Embeddings with NLP data}
This section builds on fine-tuning with a more practical approach to natural language data. We embedded the textual data into high-dimensional vectors to create an ability to compare semantic descriptions of molecules and cluster them based on those similarities. These embeddings serve a dual purpose, enabling both data visualization and acting as input to encoder-decoder structures in large language models (LLMs) like GPT and Llama. To achieve this, various embedding models were employed, including three BERT-based models and one from OpenAI.

The first model utilized was Base BERT uncased, along with Arzington BERT and Chemical BERT. Multiple embedding models allowed for the encoding of the data in different ways, capturing various aspects of its semantics. As shown in Figure 6, an example visualization of these embeddings using the BERT model, with dimensions reduced through the t-SNE algorithm (t-Distributed Stochastic Neighbor Embedding), can be observed. t-SNE is particularly valuable for visualizing high-dimensional data, such as word embeddings generated by models like BERT. BERT embeddings typically reside in a high-dimensional space (e.g., 768 dimensions), making it challenging to directly visualize and interpret relationships between words or sentences. t-SNE projects these embeddings into a lower-dimensional space, providing a 2D visual representation of semantic relationships and clusters while preserving the underlying data and relationships from the higher dimensions.

These lower-dimensional representations visually depict the semantic meanings of the descriptions relative to each other. Moreover, basic queries can be compared using cosine similarity across the entire high-dimensional vector database, enabling efficient data sorting and querying. 
\begin{figure}[htbp]
    \centering
    \includegraphics[width=\linewidth]{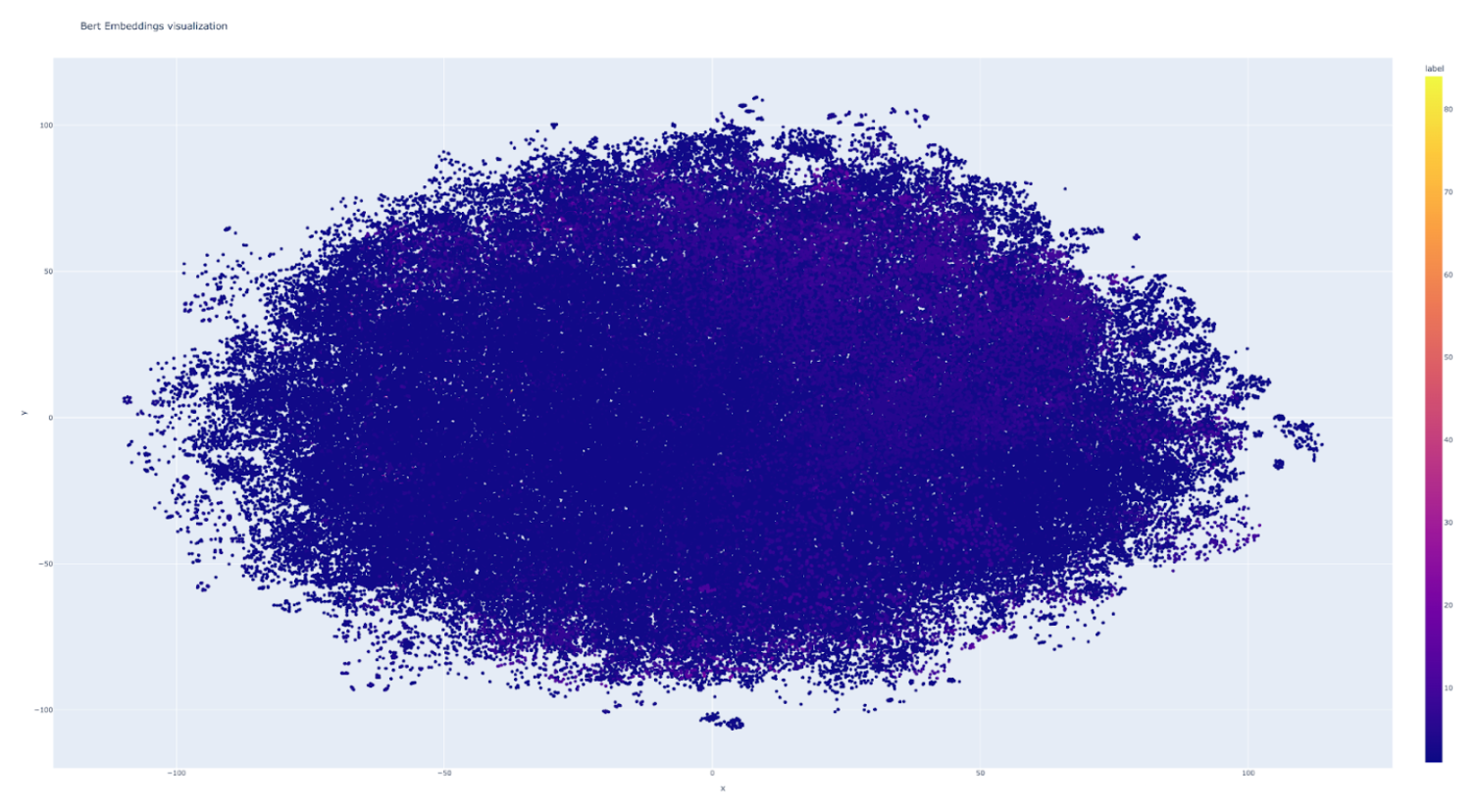}
    \caption{t-SNE Visualization of BERT Embeddings}
    \label{BERT_Embeddings.png}
\end{figure}
Fig 2 shows the t-SNE embeddings visually represented. We stored and vectorized the embedding data as "context" for an LLM, specifically GPT-3.5-turbo. The BERT embeddings served as context for the model. The BERT embeddings were saved to a vector database using DeepLake and Langchain libraries. The encoded data, combined with user prompts, resulted in human language output generated by the model, incorporating contextual information from the compound descriptions in PubChem. This approach represents a form of prompt engineering using a vast dataset of embeddings as context, demonstrating significant improvements over simple search algorithms through the vector database and the base GPT-3.5 model.

The LLM utilizing description embeddings also exhibited the ability to generalize to information not explicitly stated in the description dataset, providing additional data. This suggests that the LLM can intelligently search through the chemical dataset via natural language prompts. Running the same LLM with a more strictly curated dataset (only the larger 100,000 descriptions from the original 328,000) yielded similar results, although not statistically significant enough to warrant exclusive use of the second method. However, it did reduce the size of embeddings, which in turn reduced computational power requirements and API costs proportionally.

This method produced improved results compared to fine-tuning in being able to understand molecular properties and relationships, and it laid the foundation for comparisons with the vector search method later in the paper.

\subsection{Benchmark Data for LLM}
We created a benchmark dataset to measure the search capabilities of methods such as embedding vector search, embeddings with GPT, and a fine-tuned Llama 2 model. An ideal benchmark would consist of questions posed to the model, encompassing various properties of a given molecule. The model would then be evaluated based on the accuracy and fluency of its responses. However, the model must not possess prior knowledge of the description or question to prevent it from recognizing phrases and "cheating" to obtain answers. To overcome this challenge, descriptions of molecules were considered for scraping from Wikipedia to use as input for the model. However, simply inputting this data into the model would be problematic, as LLMs like GPT and Llama have substantial knowledge of Wikipedia from their training data, potentially recognizing phrases. To address this, summaries of the data were generated using a model, and these summaries were applied to each Wikipedia description to obtain a general summary without copying phrases directly.

This process involved using 1,000 descriptions sourced from PubChem, along with 1,000 descriptions present in PubChem but not included in the training dataset. The resulting benchmark dataset comprises 2,000 benchmark molecules, with 1,000 drawn from the embeddings and 1,000 excluded. This dataset served as the benchmark for evaluating the performance of the LLMs, primarily comparing the results of the GPT-3.5 base model and the implementation with GPT-3.5 drawing context from the vector database of embeddings defined earlier. Cosine similarity was employed as the metric for comparing model-generated outputs with desired outputs in the benchmark dataset.

\subsection{Results for LLM Model}
Using the benchmark defined previously, an evaluation of the GPT base model and the GPT model utilizing the BERT embeddings vector database was conducted. Both models underwent evaluation on the entire benchmark dataset and were queried with a specified molecule to generate a response containing the description of that molecule. Additionally, they were presented with the same prompt via prompt engineering, querying about the fundamental properties of the molecule, such as density, solubility, and appearance. This approach differed from the previous method, which queried for a specific molecule with its description provided as input, as it was deemed too variable, with multiple possible "correct" answers for a single description query, potentially leading to unwanted ambiguity in accuracy calculations. The results, evaluated using semantic similarity, are depicted in Fig 4.
\begin{figure}[htbp]
    \centering
    \includegraphics[width=\linewidth]{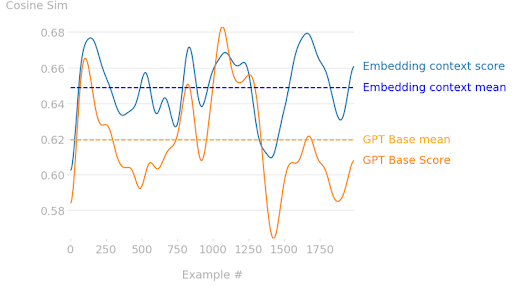}
    \caption{GPT Embeddings Context results}
    \label{GPT Embeddings Context results.png}
\end{figure}

\section{Vector Search}
\subsection{Overview}
The vector search algorithm diverges from previous methods and delves into clustering and finding relationships in non-textual data in molecules. It finds similarities between a given query molecule and the molecules in the SMILES dataset provided by PubChem. It achieves this by first clustering fingerprints of the data based on numerical measures of molecular similarity, and then returning the cluster closest to a given molecule when evaluating based on the Mahalanobis distance metric.\par Due to computational restrictions, the algorithm was only run on the first 1000 molecules from the SMILES dataset for the hyperparameter tuning and testing portions of our project. We ran the algorithm on the first 1200 molecules for clustering and cluster visualizations as pictured in our results section, as well as a visualization for the first 2000 molecules.

\subsection{Embeddings with SMILES data}
This section dives into ignoring semantic relationships between the words in each description; it does not take any textual descriptions as input. Instead, it is just given a graphical representation of SMILES data to learn emergent properties that define certain characteristics. Embeddings Using SimCLR The (SimCLR) architecture is a powerful deep-learning framework designed for generating meaningful representations from raw data. We utilized SimCLR to create embeddings of RDKit Morgan fingerprints derived from SMILES data, which encode molecular structures. SimCLR employs a Siamese network structure, where two identical subnetworks share the same weights. It learns by maximizing the similarity between positive pairs (samples from the same data point) while minimizing the similarity between negative pairs (samples from different data points). We use the cosine similarity function to do this in our case. This process encourages the network to learn high-level features that capture the intrinsic structure of the input data, making it well-suited for generating informative embeddings of molecular fingerprints. Fig 7 shows a 2D visualization of the SimCLR embeddings using the t-SNE method described earlier.

\begin{figure}[htbp]
    \centering
    \includegraphics[width=\linewidth]{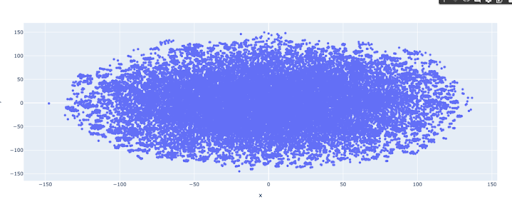}
    \caption{SimCLR Embeddings Visualization}
    \label{SimCLR Embeddings Visualization.png}
\end{figure}

These embeddings can be valuable for tasks such as molecular similarity analysis, compound screening, and drug discovery, as they encode the underlying chemical properties and relationships between molecules in a continuous vector space. Using the SimCLR architecture, we extracted the embeddings of the SMILES data for each compound within the PubChem dataset. We used a pre-trained base encoder, Resnet, as the backbone of our architecture. Another crucial aspect of the SimCLR architecture is the custom contrastive loss function, which aims to maximize the agreement between positive pairs while minimizing the agreement between negative pairs. This process tells the encoder to map similar molecules closer together in the embedding space and push dissimilar molecules apart.

\subsection{Fingerprinting and Clustering}
We use MACCS (Molecular ACCess Systems) fingerprinting, a commonly used molecular fingerprint found in the RDkit python library, in order to find similarities between given molecules. The inputs to the MACCS fingerprinting function are given by SMILES strings of given molecules. We utilize a Tanimoto similarity calculation with this metric to find similarity between two candidate molecules, which is then translated into a distance between them. This is then entered as a parameter into a matrix of given distances between pairs of molecules. \par We utilize t-SNE as a dimensionality reducer, where each row represents a given query molecule, and each column describes the Tanimoto similarity between that molecule and the others within the dataset as a feature of that molecule. We chose t-SNE because it preserves local structures within data, especially relationships between points that lie close together, making it fit for clustering tasks such as this. Our visualization, clustering, and query function use standardized (z-score normalized) forms of the data returned from the t-SNE algorithm to ensure that each feature contributes equally to similarity calculations and that outliers or large coordinates do not interfere with clustering. As an added bonus, standardization also ensures visualizations can be done with ease.\par We employ affinity propagation clustering to sort our data into groups on the basis of molecular similarity. We chose affinity propagation over other clustering algorithms due to its ability to sort without a given number of clusters as a parameter, as is the case with k-means and other clustering algorithms, thus preserving inherent patterns present in the data when visualizing and finding the closest cluster. Fig 5 visualizes the clusters of our molecules.
\begin{figure}[htbp]
    {\includegraphics[width=1.0\textwidth]{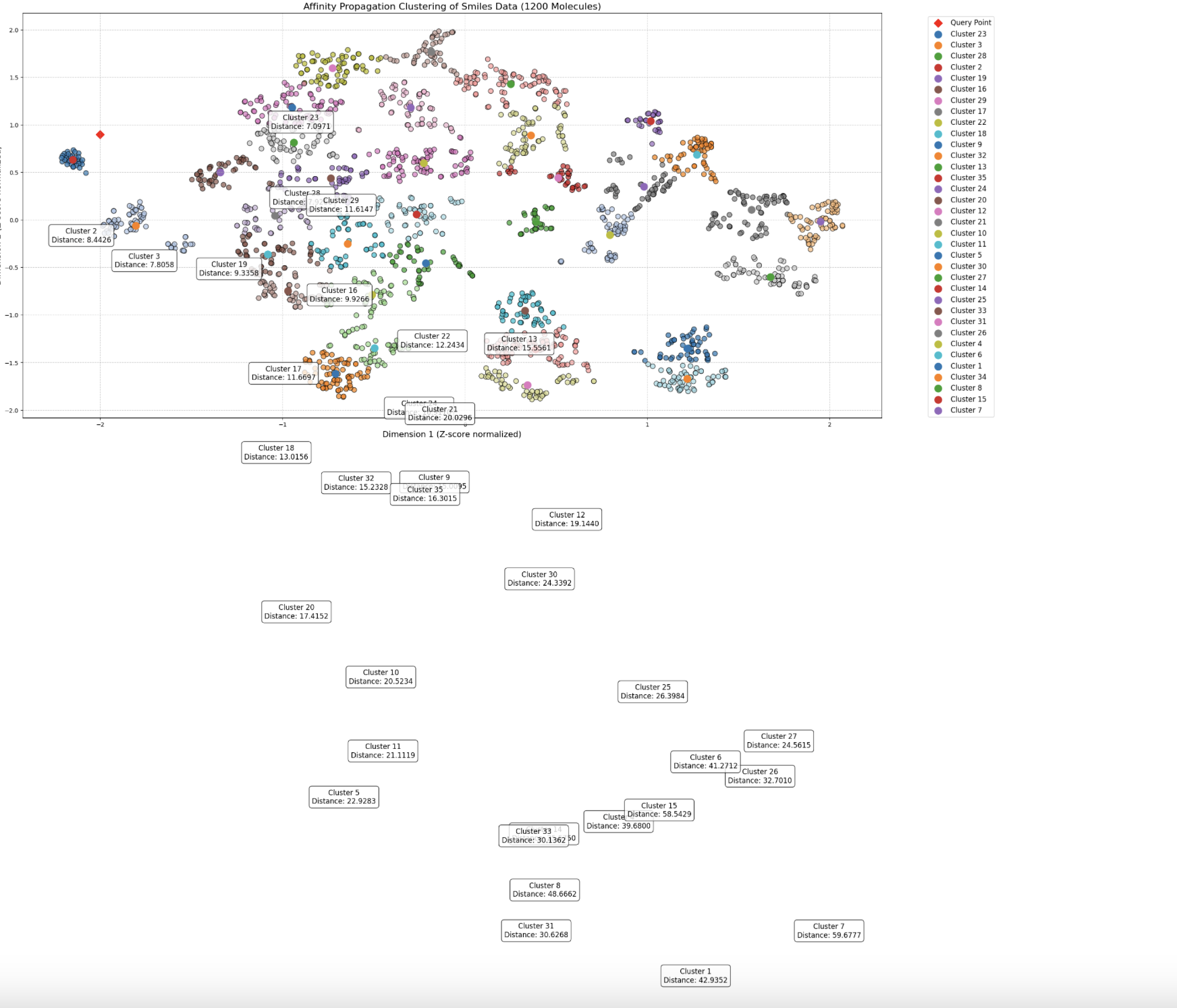} }
    \caption{Vector Search clustering}
    \label{Vector Search Clustering.png}
\end{figure}

\subsection{Query Function}
Our query function utilizes the Mahalanobis distance metric to find the closest cluster to a given query point. The Mahalanobis distance is a generalized form of the Euclidean distance metric and finds the distance between a value and a distribution of points. The Mahalanobis distance also takes into account the relationship between the parameters when looking at multivariate data, and takes into account the variability within the data to return the number of standard deviations away from the mean at which a given data point is located.

We chose the Mahalanobis distance metric for these particular properties. The distances produced by the compute\_tanimoto\_distances function within our code are the basis of molecules related to the query and do not take into account the fact that each of the parameter molecules may be related to each other as well. The usage of the Mahalanobis distance metric accounted for this factor and returned a statistically significant result of the similarity between a given query point and the clusters.

\subsection{Tanimoto Accuracy Calculation}
We measure the accuracy of our clustering model by implementing Tanimoto accuracy calculation between the molecules of each given cluster. Using the query function, we input precise values to get specific clusters as the closest cluster, and then implement a Tanimoto accuracy calculation upon them, using MACCS fingerprinting. In addition, we calculate accuracy via summary statistics, as pictured in the results section.

\subsection{Time Complexity for Vector Search}

The visualization which showcases all the clusters and where they are centered took about 8-10 minutes to retrieve. The data preprocessing, including the one-hot encoding, and computing distances, can be computationally intensive especially when working with a large dataset of 1000-2000 long SMILE strings. Along with this, the t-SNE used for dimensionality reduction can be time-consuming with large datasets due to the pairwise computations and optimization it involves to find the best low-dimensional representation. 

The affinity propagation clustering algorithm applied to the reduced-dimensional data is computationally expensive, especially since the algorithm needs to iterate many times to converge. The visualization, although may not be computationally intensive by itself, can add to the overall runtime. 

The visualizations represent the query points and the summary statistics took about 1 minute to retrieve. The time to retrieve this is relatively faster due to the smaller size of the input data. This is visualizing the Tanimoto coefficients for a subset of molecules from the closest cluster, which is expected to be much smaller than the original dataset. Additionally, generating a bar chart is less computationally expensive compared to dimensionality reduction and clustering. Lastly, these visualizations require no iterative algorithms or complex calculations, which reduces its runtime.

\subsection{Vector Search Results}
Our method describes how clustering algorithms, when modified with chemical fingerprinting, can describe similarity between molecules to a certain extent. Though this does not describe emergent properties of understanding molecular design and properties by a model, it analyzes molecular similarity from another angle, particularly the usefulness of clustering algorithms when querying based on similarity. The algorithm uses SMILES data to perform similarity search to get an idea of what molecular similarity might look like based on related molecules. This is an alternative method to search for molecular properties; rather than using prompt-based queries, we use Tanimoto similarities as distances to cluster molecules, while leaving open the possibility for custom embeddings of the data to improve upon our accuracy. Given a numerical representation of a molecule within our framework, the vector search model is then able to return the cluster of molecules that describes the molecules most similar to it, an essential component of understanding the underlying properties that make molecules similar.

\begin{figure}[!tbp]
  \centering
  \begin{minipage}[b]{0.45\textwidth}
    \includegraphics[width=\textwidth]{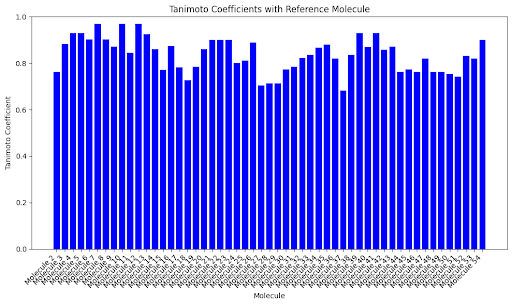}
    \caption{Tanimoto Coefficients with Reference Molecule}
  \end{minipage}
  \hfill
  \begin{minipage}[b]{0.45\textwidth}
    \includegraphics[width=\textwidth]{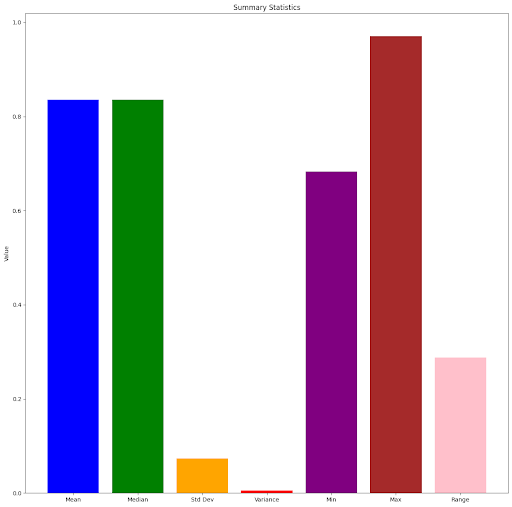}
    \caption{Summary Statistics}
  \end{minipage}
  
\end{figure}

Numerically, our findings satisfactorily met the key threshold value of 0.6. For most of the representative clusters below, our model met the threshold. The results were more highly pronounced with concentrated clusters, achieving the highest accuracy with the cluster centered at [-2.0,1.0]. 

For a majority of non-representative clusters, we met the threshold of 0.55. Below we have pictured summary statistics and representative clusters for a variety of point queries to our model.


\begin{thebibliography}{00}

\bibitem{b1} Bran, A. M., Cox, S., White, A. D., \& Schwaller, P. (2023, June 21). ChemCrow: Augmenting large-language models with Chemistry Tools. arXiv.org. https://arxiv.org/abs/2304.05376

\bibitem{b2} Chen, T., Kornblith, S., Norouzi, M., \& Hinton, G. (2020, July 1). A simple framework for contrastive learning of visual representations. arXiv.org. https://arxiv.org/abs/2002.05709

\bibitem{b3} Choy, C. T., Wong, C. H., \& Chan, S. L. (2019). Embedding of genes using cancer gene expression data: Biological relevance and potential application on Biomarker Discovery. Frontiers in Genetics, 9. https://doi.org/10.3389/fgene.2018.00682

\bibitem{b4} Flam-Shepherd, D., Zhu, K., \& Aspuru-Guzik, A. (2022). Language models can learn complex molecular distributions. Nature Communications, 13(1). https://doi.org/10.1038/s41467-022-30839-x

\bibitem{b5} Grisoni, F. (2023). Chemical language models for de Novo Drug Design: Challenges and opportunities. Current Opinion in Structural Biology, 79, 102527. https://doi.org/10.1016/j.sbi.2023.102527

\bibitem{b6} Guo, H., Huo, M., Zhang, R., \& Xie, P. (2023). ProteinChat: Towards achieving chatgpt-like functionalities on protein 3D structures. Tech Rxiv IEEE. https://doi.org/10.36227/techrxiv.23120606

\bibitem{b7} Hao, Z., Lu, C., Huang, Z., Wang, H., Hu, Z., Liu, Q., Chen, E., \& Lee, C. (2020). ASGN: An active semi-supervised Graph Neural Network for Molecular Property Prediction. Proceedings of the 26th ACM SIGKDD International Conference on Knowledge Discovery \& Data Mining. https://doi.org/10.1145/3394486.3403117

\bibitem{b8} Hu, S., Chen, P., Gu, P., \& Wang, B. (2022). A deep learning-based chemical system for QSAR prediction. IEEE Journal of Biomedical and Health Informatics, 24(10), 3020–3028. https://doi.org/10.1109/jbhi.2020.2977009

\bibitem{b9} Johnson, J., Douze, M., \& Jegou, H. (2021). Billion-scale similarity search with GPUs. IEEE Transactions on Big Data, 7(3), 535–547. https://doi.org/10.1109/tbdata.2019.2921572

\bibitem{b10} Kulmanov, M., Smaili, F. Z., Gao, X., \& Hoehndorf, R. (2020). Semantic similarity and Machine Learning with ontologies. Briefings in Bioinformatics, 22(4). https://doi.org/10.1093/bib/bbaa199

\bibitem{b11} Liu, H., Xiong, J., Zhang, N., \& Zhong, J. (2022). Text modality enhanced based deep hashing for multi-label cross-modal retrieval. 2022 14th International Conference on Advanced Computational Intelligence (ICACI). https://doi.org/10.1109/icaci55529.2022.9837775

\bibitem{b12} Maaten, L. van der, \& Hinton, G. (1970, January 1). Visualizing data using T-Sne. JMLR. https://www.jmlr.org/papers/v9/vandermaaten08a.html

\bibitem{b13} Mardt, A., Pasquali, L., Wu, H., \& Noé, F. (2018). Vampnets for deep learning of molecular kinetics. Nature Communications, 9(1). https://doi.org/10.1038/s41467-017-02388-1

\bibitem{b14} Mohamed, S. K., Nounu, A., \& Nováček, V. (2020). Biological applications of knowledge graph embedding models. Briefings in Bioinformatics, 22(2), 1679–1693. https://doi.org/10.1093/bib/bbaa012

\bibitem{b15} Nasser, M., Yusof, U. K., \& Salim, N. (2023). Deep learning based methods for molecular similarity searching: A systematic review. Processes, 11(5), 1340. https://doi.org/10.3390/pr11051340

\bibitem{b16} Niazi, S. K., \& Mariam, Z. (2023). Recent advances in machine learning-based chemoinformatics: A comprehensive review. MDPI. https://doi.org/10.20944/preprints202306.0803.v1

\bibitem{b17} Pan, J. (2023). Large language model for Molecular Chemistry. Nature Computational Science, 3(1), 5–5. https://doi.org/10.1038/s43588-023-00399-1

\bibitem{b18} Rong, Y., Bian, Y., Xu, T., Xie, W., Wei, Y., Huang, W., \& Huang, J. (2020, October 29). Self-supervised graph transformer on large-scale molecular data. arXiv.org. https://doi.org/10.48550/arXiv.2007.02835

\bibitem{b19} Ross, J., Belgodere, B., Chenthamarakshan, V., Padhi, I., Mroueh, Y., \& Das, P. (2022). Molformer: Large scale chemical language representations capture molecular structure and properties. Nature Machine Intelligence. https://doi.org/10.21203/rs.3.rs-1570270/v1

\bibitem{b20} Saxena, D., Sharma, A., Siddiqui, M. H., \& Kumar, R. (2019). Blood brain barrier permeability prediction using machine learning techniques: An update. Current Pharmaceutical Biotechnology, 20(14), 1163–1171. https://doi.org/10.2174/1389201020666190821145346

\bibitem{b21} Siu, S. C. (2023). Chatgpt and GPT-4 for professional translators: Exploring the potential of large language models in translation. SSRN Electronic Journal. https://doi.org/10.2139/ssrn.4448091

\bibitem{b22} Tang, Q., Nie, F., Zhao, Q., \& Chen, W. (2022). A merged molecular representation deep learning method for blood–brain barrier permeability prediction. Briefings in Bioinformatics, 23(5). https://doi.org/10.1093/bib/bbac357

\bibitem{b23} Wang, Yanbin, You, Z.-H., Yang, S., Li, X., Jiang, T.-H., \& Zhou, X. (2019). A high efficient biological language model for predicting protein–protein interactions. Cells, 8(2), 122. https://doi.org/10.3390/cells8020122

\bibitem{b24} Wang, Yuyang, Wang, J., Cao, Z., \& Barati Farimani, A. (2022). Molecular contrastive learning of representations via Graph Neural Networks. Nature Machine Intelligence, 4(3), 279–287. https://doi.org/10.1038/s42256-022-00447-x

\bibitem{b25} Wu, Z., Ramsundar, B., Feinberg, E. N., Gomes, J., Geniesse, C., Pappu, A. S., Leswing, K., \& Pande, V. (2017). MoleculeNet: A benchmark for Molecular Machine Learning. Chemical Science, 9(2), 513–530. https://doi.org/10.1039/c7sc02664a

\bibitem{b26} Xu, Z., Wang, S., Zhu, F., \& Huang, J. (2017). Seq2seq Fingerprint: An Unsupervised Deep Molecular Embedding for Drug Discovery. Proceedings of the 8th ACM International Conference on Bioinformatics, Computational Biology, and Health Informatics. https://doi.org/10.1145/3107411.3107424

\bibitem{b27} Zhang, X.-C., Wu, C.-K., Yi, J.-C., Zeng, X.-X., Yang, C.-Q., Lu, A.-P., Hou, T.-J., \& Cao, D.-S. (2022). Pushing the boundaries of molecular property prediction for drug discovery with multitask learning Bert Enhanced by smiles enumeration. Research, 2022. https://doi.org/10.34133/research.0004

\bibitem{b28} Zhao, S., Zhang, J., \& Nie, Z. (2023, June 7). Large-scale cell representation learning via divide-and-conquer Contrastive Learning. arXiv.org. https://arxiv.org/abs/2306.04371

\end{thebibliography}
\end{document}